\documentclass[aps,prl,twocolumn,showpacs]{revtex4}  
\usepackage{epsfig}
\begin{document}
 \title{Nonexistence of intrinsic  spin currents}
 \author{Alexander  Khaetskii}
\address{Physics Department and Center for NanoScience, LMU
M\"unchen, D-80333, Germany}

\date{\today}

\draft

\begin{abstract}
We have described the electron spin dynamics in the presence of the spin-orbit interaction and disorder using the spin-density matrix method. 
We showed that in the Born approximation in the scattering amplitude  
 the spin current is zero for an arbitrary ratio of  the spin-orbit splitting
and  the scattering rate.
 Various types of the  disorder potential are studied. We argue that the bulk spin current has always an {\it extrinsic} nature and  depends  explicitely on scattering by impurities since it appears only beyond the Born approximation in the scattering amplitude. 
\end{abstract}

\maketitle

Spin-orbit coupling brings about a number of  interesting effects, one of which   is generation of a spin flux in the plane perpendicular to the charge current direction. This phenomenon occurs in the paramagnetic system and is very well known for quite a long time, see Ref.\cite{Dyakonov}, where the Yafet-Elliott spin-orbit mechanism was considered.  It is a consequence of the fact that in the presense of spin-orbit coupling the scattering by impurities has  an asymmetric character (the Mott effect) \cite{Landau}.  Spins with up-orientation are scattered preferably to the right and with down-orientation - to the left. This phenomenon exists only beyond the Born approximation in the scattering amplitude and leads to an accumulation of the spin density near the sample surface \cite{Dyakonov}. Mutual transformation of the current and spin fluxes leads also to the renormalization of the electrical conductivity of the system, see Ref.\cite{Dyakonov1}, where the case of 3D holes described by the Luttinger Hamiltonian  was considered.  
 It has been recently claimed \cite{MacDonald,Zhang} that an analogous phenomenon can exist even without scattering by impurities, i.e. in the ballistic regime, the corresponding contribution being called "intrinsic" or dissipationless. Later several papers appeared where the effect of scattering  by impurities  was taken into account \cite{Loss,Burkov,Bauer,Dimitrova}  with a range of totally different results.
 This was done by the Kubo formula. 
We solve this problem using the  well known method of a spin-density matrix \cite{Dyakonov1}. We argue that {\it intrinsic} spin current in the bulk  cannot exist.
Any spin current must be due to asymmetric scattering by impurities (the Mott effect). Since this phenomenon occurs only
 beyond the Born approximation in the scattering amplitude, this contribution is explicitely dependent on the impurity scattering and leads to the well known {\it extrinsic} contribution. \cite{Dyakonov,Dyakonov1}. In particular, there cannot exist the universal value of the spin current since it does  not depend on the scattering properties. In the Born approximation (when the scattering amplitude has additional symmetry propeties, see below) the Mott effect is absent, and 
the spin current is zero for an arbitrary value of $\Delta \tau$, where $\Delta$ is the spin splitting of the electron 
spectrum and $\tau$ is the transport scattering time. We have shown this by  exact calculations for the case of the Rashba Hamiltonian. \cite{Luttinger}
Thus the correct terminology, i.e. intrinsic or  extrinsic contribution, should be used in accordance with  the strength of the scattering (Born or beyond) rather then the  presence or absence of the impurity scattering itself. Therefore, certain care should be taken when trying to check by ``exact''
 numerical diagonalization the robustness of the "intrinsic" value with respect to the disorder since strong scattering inevitably generates an  {\it extrinsic} contribution to the spin current. 
\par
The Hamiltonian of the problem is
\begin{equation}
\hat {\cal H({\bf p})}=\frac{p^2}{2m}+ \frac{\alpha}{2}\vec{\sigma}\cdot\vec{
\Omega}({\bf p}),
\,\,\,  \epsilon_M(p)= \frac{p^2}{2m}+ M\alpha p,
\end{equation}
where  $\vec{\Omega}({\bf p})=[{\bf n}\cdot {\bf p}]$, ${\bf n}$ is the unit vector normal to the 2D plane (z-axis), $\epsilon_M(p)$ are the eigenvalues, 
$ M=\pm 1/2$ are the helicity values. The eigenfunctions are
$$
\chi_{M{\bf p}}= \sum_{\mu=\pm 1/2} D^{(1/2)}_{\mu M}(\vec{\Omega})u_{\mu}=
 \sum_{\mu=\pm 1/2}e^{-i\mu (\phi-\pi/2)}d^{(1/2)}_{\mu M}(\frac{\pi}{2})u_{\mu},
$$
where $D^{(1/2)}_{\mu M}(\vec{\Omega})$ is the rotation matrix \cite{Landau}, 
 $\phi$  the angle of ${\bf p}$, and
$u_{\mu}$  the eigenfunction of the $\hat \sigma_z$ operator.
\par
{\it Spin current, kinetic equation}.
We will calculate the $q_{yz}$ component of the spin current. This quantity is zero in the thermodynamic limit \cite{Rashba} and defined as
\begin{widetext}
\begin{equation}
q_{yz}=Tr \int \frac{d^2p}{(2\pi)^2} \hat f({\bf p})\frac{1}{2} (\hat S_z\hat V_y +
\hat V_y \hat S_z)= Tr \int \frac{d^2p}{(2\pi)^2} 
\frac{p_y}{m}\hat f({\bf p})\hat S_z \propto (f_{+-} + f_{-+}).
\label{q}
\end{equation}
\end{widetext}
Here $\hat f({\bf p})$ is the spin density matrix \cite{Wigner}, $\hat V_y$ the $y$-component of the velocity operator and $\hat S_z=(1/2) \hat \sigma_z$  the spin operator. The last expression in  Eq.(\ref{q})
 is given in the helicity basis.
 The general expression for the quantum kinetic equation in the case of spin-orbit interaction, when the Hamiltonian and the Wigner distribution function are matrices over the spin indexes, was derived in Ref.\cite{Dyakonov1}. When there is a magnetic field or some inhomogeneity in the problem, the field term and the gradient term must be symmetrized since the velocity operator is also a matrix. In our  case  when we deal only with the electric field which is constant in space this equation is simple and reads
\begin{equation}
\frac{\partial \hat f({\bf p})}{\partial t} + e{\bf E}\frac{\partial \hat f}{\partial {\bf p}} + \frac{i}{\hbar}[\hat {\cal H({\bf p})}, \hat f] = St\{ \hat f({\bf p})\}
\label{kinetic}
\end{equation} 
The last term on the left hand side is a commutator and the expression for the collision term is given below. 
 Now we write Eq.(\ref{kinetic}) in the helicity basis where the Hamiltonian  is diagonal. While doing that, we should take into account the fact that eigenfunctions  $\chi_{M{\bf p}}$ depend on the direction of the momentum ${\bf p}$, thus the matrix elements of the derivative $\partial \hat f/\partial {\bf p}$ in this basis do not coincide with the quantities $\partial  f_{MM'}/\partial {\bf p}$
$$
\left(\frac{\partial \hat f}{\partial {\bf p}}\right)_{MM'} =  
\frac{\partial  f_{MM'}}{\partial {\bf p}} - \frac{i}{\hbar} [\hat {\bf a}, \hat f]_{MM'}; \,\, {\bf a}_{MM'} = i \hbar \chi^{\star}_{M{\bf p}}\frac{\partial \chi_{M'{\bf p}}}{\partial {\bf p}}.
$$
 We see that there  appears  the commutator of the vector matrix $\hat {\bf a}$ with $\hat f$.
Thus for Eq.(\ref{kinetic}) in the linear response regime (${\bf E} \parallel x$)  we obtain 
\begin{widetext}
\begin{equation}
eE \cos\phi \frac{\partial f^{(0)}_{MM}}{\partial p}
\delta_{MM'}-\frac{i}{2}\frac{\sin \phi}{p}eE ( f^{(0)}_{M'M'}(p)- f^{(0)}_{MM}(p) ) +
\frac{i}{\hbar}(\epsilon_M(p)-\epsilon_{M'}(p))f_{MM'}(p)= St (\hat f(p))_{MM'}
\label{kinetic1}
\end{equation}
\end{widetext}
Here $f^{(0)}_{MM}(p)$ is the equilibrium Fermi function corresponding to the helicity value $M$.
The collision term was derived in many papers, for the refs. see \cite{Dyakonov1,Koshelev}, and in the helicity basis has the form
\begin{widetext}
\begin{eqnarray}
St (\hat f({\bf p}))_{MM'}=
 \int\frac{d^2 {\bf p}_1}{(2\pi \hbar)^2} \sum_{M_1M'_1}
\{[\delta(\epsilon_{M_1}(p_1)-\epsilon_{M}(p)) +\delta(\epsilon_{M'_1}(p_1)-\epsilon_{M'}(p))]K^{MM'}_{M_1M'_1}(\omega_{{\bf p}{\bf p}_1}) \cdot f_{M_1M'_1}
({\bf p}_1)- \nonumber \\
 -\delta(\epsilon_{M_1}(p)-\epsilon_{M'_1}(p_1))[K^{MM_1}_{M'_1 M'_1 }(\omega_{{\bf p}{\bf p}_1}) \cdot f_{M_1M'}({\bf p}) + f_{MM_1}({\bf p})\cdot K^{M_1M'}
_{M'_1 M'_1}(\omega_{{\bf p}{\bf p}_1}) ] \}, 
\label{St}
\end{eqnarray} 
\end{widetext}
where the kernel in the Born approximation in the scattering amplitude is: 
\begin{equation}
K^{MM'}_{M_1M'_1}(\omega_{{\bf p}{\bf p}_1})= \frac{\pi}{\hbar}
|U({\bf p} -{\bf p_1})|^2 \cdot N  D^{(1/2)}_{MM_1}(\omega_{{\bf p}{\bf p}_1})
D^{(1/2)\star}_{M'M'_1}(\omega_{{\bf p}{\bf p}_1}).
\label{kernel}
\end{equation}
Here $N$ is the 2D impurity density, $U({\bf p} -{\bf p_1})$ is the Fourier component of the impurity potential.  $D^{(1/2)}_{MM_1}(\omega_{{\bf p}{\bf p}_1})$ depends only on the scattering angle $\theta =\phi -\phi_1$. Diagonal components $D^{(1/2)}_{1/2,1/2}=D^{(1/2)}_{-1/2,-1/2}= \cos (\theta/2)$, 
and $D^{(1/2)}_{1/2,-1/2}=D^{(1/2)}_{-1/2,1/2}= -i \sin (\theta/2)$. 
The Born scattering amplitude is given by
\begin{equation}
F^{M{\bf p}}_{M_1{\bf p}_1} \propto D^{(1/2)}_{MM_1}(\omega_{{\bf p}{\bf p}_1})U({\bf p} -{\bf p_1}); \,\,\, F^{M{\bf p}}_{M_1{\bf p}_1}= (F^{M_1{\bf p}_1}_{M{\bf p}})^{\star}.
\label{Born}
\end{equation}
 The additional symmetry property indicated here exists only in the Born approximation. \cite{Landau}
\par
{\it Smooth scattering potential.}
 First  consider the mathematically simple case of a smooth scattering potential when  the interband transitions are suppressed  which is realized at $m\alpha R/\hbar \gg 1$, $R$ being  the radius of impurity.
 Then from  Eqs.(\ref{kinetic1},\ref{St}) we obtain
\begin{eqnarray}
eE\frac{\partial f^{(0)}_+}{\partial p}&=& \frac{2ap}{V_+}f_{++} + 
\frac{bp}{V_+}(f_{+-} - f_{-+}), 
\label{sm1} \\
eE\frac{\partial f^{(0)}_-}{\partial p}&=& \frac{2ap}{V_-}f_{--}- 
\frac{bp}{V_-}(f_{+-} -f_{-+}),  
\label{sm2} 
\end{eqnarray}
\begin{widetext}
\begin{eqnarray}
\frac{i}{2}\frac{eE}{p}(f^0_+ - f^0_-) + \frac{i}{\hbar}(\epsilon_+ -\epsilon_- )f_{+-} &=& c(\frac{p}{V_+}f_{++}- 
\frac{p}{V_-}f_{--})+ pd (\frac{1}{V_+} +\frac{1}{V_-})f_{+-},  
\label{sm3} \\
\frac{-i}{2}\frac{eE}{p}(f^0_+ - f^0_-) -\frac{i}{\hbar}(\epsilon_+ -\epsilon_- )f_{-+} &=& -c(\frac{p}{V_+}f_{++}- 
\frac{p}{V_-}f_{--})+ pd (\frac{1}{V_+} +\frac{1}{V_-})f_{-+},
 \label{sm4}
\end{eqnarray} 
 \end{widetext}
where $d=a, b=-c, c=-ia, a= -\frac{1}{2}\int d\theta/(2\pi) W(\theta) \sin^2 \theta$, quantity $-2ap/V_+$ is equal to the inverse  transport scattering time $\tau$, $W(\theta)= N \cdot |U({\bf p} -{\bf p_1})|^2/2\hbar^3$.
$f^0_+(p),f^0_-(p)$ are the equilibrium Fermi functions which correspond to the helicity 
$\pm$, $V_{\pm}(p)=p/m \pm \alpha/2$ are the velocity values for a given $p$ for $\pm$ bands.
 The expressions for the coefficients $a,b,c,d$  are exact but  should be used here for $\theta \ll 1$ since we consider  small-angle scattering. In  Eqs.(\ref{sm1}-\ref{sm4}) the quantities $f_{++}(p),f_{+-}(p),f_{-+}(p),f_{--}(p)$ depend only on the modulus of ${\bf p}$. In deriving these Eqs. we used the following angular dependences of the components of the matrix $\hat f({\bf p})$: 
$f_{++}({\bf p}),f_{--}({\bf p}) \propto \cos\phi$ and $f_{+-}({\bf p}), f_{-+}({\bf p})\propto \sin\phi$.
Besides, we used the symmetry properties of the matrix $K(\theta)$:
\begin{equation}
 K^{++}_{++} = K^{--}_{--}, \,\, K^{++}_{-+} = K^{-+}_{++}, \,\, K^{+-}_{++} = - K^{-+}_{++},
\label{symmetry1}
\end{equation}
which  can be easily proved from Eqs.(\ref{kernel},\ref{Born}).
The quantities entering Eqs.(\ref{sm1}-\ref{sm4}) have the following relations to the average spin components: 
\begin{widetext}
\begin{equation}
<S_z> \propto (f_{+-}+f_{-+}),\,\,\,  <\vec{S}\cdot\vec{p}> \propto 
 (f_{+-}-f_{-+}), \,\,\, <\vec{S}\cdot\vec{\Omega}> \propto 
 (f_{++}-f_{--}).
\label{relation}
\end{equation}
\end{widetext}
The last quantity exists even in the thermodynamic limit (for a given momentum ${\bf p}$). 
 From the above equations  for the quantity of interest we find
\begin{widetext}
\begin{equation}
eE( \frac{\partial f^0_{+}}{\partial p}-\frac{\partial f^0_{-}}{\partial p})
+ \frac{eE}{p}(f^0_+ - f^0_-)= -\frac{1}{\hbar} (\epsilon_+ -\epsilon_- )
(f_{+-}+f_{-+}).
\label{result1}
\end{equation}
\end{widetext}
This equation is exact for an arbitrary value of $\Delta \tau$, $\Delta= (\epsilon_+ -\epsilon_-)=\alpha p$. Hence, $q_{yz}=0$, i.e. the spin current is zero. Note that Eq.(\ref{result1}) has a clear physical meaning. The second term on the left hand side was taken into account before \cite{MacDonald} and describes the appearence in the electric field of the z-component of the spin due to the angular dependence of the wave functions. Exactly this term gives the contribution $e/8\pi$ after integration in Eq.(\ref{q}). However, the first term in Eq.(\ref{result1}) describes the change in the distribution functions due to the accelaration along the electric field and cancels  exactly  the contribution of the second term after integration in Eq.(\ref{q}). \cite{Dyakonov2}
Note that in Eqs.(\ref{sm1},\ref{sm2}) the scattering admixes only the  component $(f_{+-} - f_{-+})$. This is the direct consequence of the Born approximation and the symmetry properties, Eq.(\ref{symmetry1}). Beyond the Born approximation the quantity 
$(K^{+-}_{++} +  K^{-+}_{++})$ is not zero. Exactly this quantity is responsible for the generation of the spin flux due to the scattering when the particle flux flows in the sample, see Ref.\cite{Dyakonov1}. When this quantity is not zero, in Eqs.(\ref{sm1},\ref{sm2}) there  appears the term proportional to $(f_{+-} + f_{-+})$  which means the appearence of $q_{yz}$ due to the Mott effect  when
the  current flows in the x-direction. 

\par
{\it $\delta$-scattering potential.} Here  we consider the case of a short range scattering potential when $W(\theta)=W_0$ (constant). Then interband transitions are allowed and from Eqs.(\ref{kinetic1},{\ref{St}})  we obtain  a  system of coupled equations similar to Eqs.(\ref{sm1}-\ref{sm4}) where now the components of spin-density matrix for the values of $p_{\pm}=p\pm m\alpha$ appear (see Fig.1). To simplify the presentation we will consider  only the limiting cases of large and small  $\Delta \tau$.
When $\hbar/\tau \ll \Delta$ we find  that
$(f_{+-} -f_{-+})/(f_{+-} + f_{-+}) \simeq (\hbar/\tau)/\Delta \ll 1$. 
Neglecting everywhere the $(f_{+-} -f_{-+})$ components,
 for the z-component of the spin we obtain 
\begin{figure}
\vspace{0.cm}\narrowtext
{\epsfxsize=7.5cm
\centerline{{\epsfbox{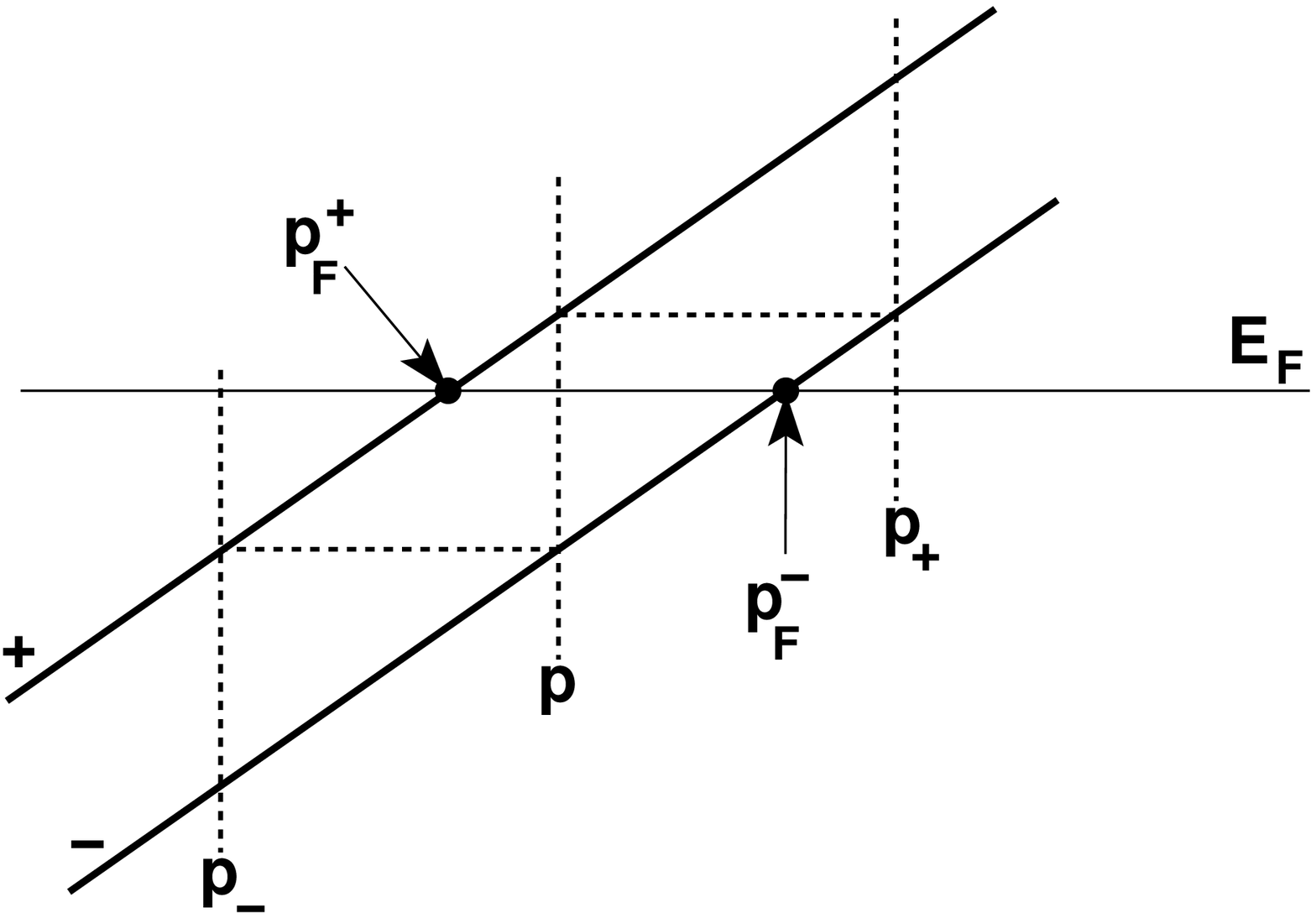}}}}
\caption{Schematics of the $\pm$ energy bands. Momenta $p, p_{\pm}, p^{\pm}_F$ are shown, see the text.}
\label{SpinZero}
\end{figure}
\begin{eqnarray}
\frac{ieE}{p}(f^0_+(p) - f^0_-(p))+ \frac{i}{\hbar} (\epsilon_+(p)&-&\epsilon_-(p) )\left (f_{+-}(p)+f_{-+}(p)\right) \nonumber \\
=  2c X_p + 2c Y_p,  \nonumber
\end{eqnarray}
where 
\begin{equation}
X_p=\frac{pf_{++}(p)}{V_+}- \frac{pf_{--}(p)}{V_-}, \,\,\,  
Y_p=\frac{p_-}{V_-}f_{++}(p_-)- \frac{p_+}{V_+}f_{--}(p_+).
\label{intermid}
\end{equation}
Note that in Eq.(\ref{intermid}) the velocities $V_{\pm}(p)$ enter at the  momentum $p$ since for horizontal 
transition the velocity is conserved. For the $f_{++}, f_{--}$
components we have the equations
\begin{eqnarray}
E_+(p)&=& \frac{2ap}{V_+}f_{++}(p)+ 
\frac{2a_1p_+}{V_+}(f_{--}(p_+)+2f_{++}(p)), \nonumber \\
E_-(p)&=& \frac{2ap}{V_-}f_{--}(p)+
\frac{2a_1p_-}{V_-}(f_{++}(p_-)+2f_{--}(p)),  \nonumber 
\end{eqnarray}
where we introduced the notations: $E_+(p)=eE \partial f^0_{+}(p)/\partial p$, $E_-(p)=eE \partial f^0_{-}(p)/\partial p$ and
 again  $V_{\pm}(p)$ enter at the  momentum $p$.
 $a_1=\frac{W_0}{2} \int d\theta/(2\pi)(1-\cos\theta)\cos\theta=-W_0/4$.
Writing these Eqs. for the momenta values $p_{\pm}$, using $a_1=a$ and the relations $V_+(p)=V_-(p_+)$, $V_+(p_-)=V_-(p)$, $ f^0_{+}(p)=
 f^0_{-}(p_+)$, $f^0_{+}(p_-)= f^0_{-}(p)$, we obtain 
\begin{equation}
p_+ f_{++}(p)=pf_{--}(p_+), \,\,\,  p_- f_{--}(p)=pf_{++}(p_-).
\label{2a}
\end{equation}
Then the solution is 
\begin{equation}
f_{++}(p)= E_+(p)
\frac{pV_+}{2a}\frac{1}{(p+p_+)^2},
\label{3}
\end{equation}
and the expression for $f_{--}(p)$ is obtained from Eq.(\ref{3}) by replacing $+$ by $-$. Using Eqs.(\ref{3},\ref{2a}) we can calculate quantities $X_p, Y_p$ and with the use of $c=-ia$ finally obtain 
\begin{widetext}
\begin{equation}
\frac{eE}{p}(f^0_+ - f^0_-)-  
m\alpha \left( \frac{E_+(p)}{2p+m\alpha} + \frac{E_-(p)}{2p-m\alpha}\right )
= -\frac{1}{\hbar} (\epsilon_+ -\epsilon_- )
(f_{+-}+f_{-+})
\label{result2}
\end{equation}
\end{widetext}
Here the ratio between $m\alpha$ and $p$ is arbitrary, the  only  restriction is that the position of the Fermi level should correspond to the values of $p$ where the above mentioned equalities between the velocities are still valid, see Fig.1. 
Integrating in Eq.(\ref{q}) between the points $p_F^{\pm}= \mp m\alpha/2 + \sqrt{p_F^2 + (m\alpha)^2/4}$, see Fig.1, we again obtain  $q_{yz}=0$.
In the opposite case $\Delta =0$ we immediately see from Eqs.(\ref{kinetic1},\ref{St}) that $(f_{+-}+f_{-+})= 0$ and the spin current is zero.
\par
In conclusion, using the spin-density matrix method for the case of the Rashba Hamiltonian we showed that  within the Born approximation in the scattering amplitude the intrinsic spin current is zero for an arbitrary ratio of spin splitting and the impurity scattering rate. We argue that the spin current appears only beyond the Born approximation, depends explicitely on the scattering and corresponds to the well known {\it extrinsic}  spin currents \cite{Dyakonov,Dyakonov1}. 
\par
After this work had been completed I became aware of  recent work \cite{Halperin}. It is not clear to me to what extent the authors generalize their conclusion about the absence of the spin currents in the bulk. Again, my opinion is that {\it extrinsic} dc spin currents can flow even in the bulk  \cite{Dyakonov,Dyakonov1}. 
\par
I am grateful to M.I. D'yakonov, L. Glazman and E.I. Rashba for fruitful discussions and also to the participants of the PASPSIII conference for the interest to my work.



\begin{references}

\bibitem{Dyakonov} 
M.I. D'yakonov and V.I. Perel, Physics Letters {\bf 35A}, 459 (1971).

\bibitem{Landau}
L.D. Landau, E.M. Lifshitz, Quantum Mechanics, Addison-Wesley, 1968.   

\bibitem{Dyakonov1} 
M.I. D'yakonov and A.V. Khaetskii, Sov. Phys. JETP {\bf 59}, 1072 (1984).

\bibitem{MacDonald}
J. Sinova et al., cond-mat/0307663, PRL {\bf 92}, 126603  (2004). 

\bibitem{Zhang} 
S. Murakami, N. Nagaosa, and S.-C. Zhang, Science {\bf 301}, 1348 (2003).

\bibitem{Loss}
J. Schliemann and D. Loss, cond-mat/0310108; PRB {\bf 69}, 165315 (2004). 

\bibitem{Burkov}
A.A. Burkov, A.H. MacDonald, cond-mat/0311328.

\bibitem{Bauer}
J. Inoue, G.E.W. Bauer, and L.W. Molenkamp,  cond-mat/0402442.
I believe that the statement made by the authors about reappearence of the universal value of  the spin current for a smooth scattering potential is incorrect.

\bibitem{Dimitrova}
O.V.  Dimitrova, cond-mat/0405339. 

\bibitem{Luttinger}
We have done similar calculations for the holes described by the Luttinger Hamiltonian, A. Khaetskii, unpublished. 
The calculations are done within the Born approximation for the collision integral when the {\it extrinsic} spin current is zero \cite{Dyakonov1}.
 These calculations show that the conclusion of the paper S. Murakami, cond-mat/0405001 about the complete absence  of the corrections to the intrinsic value of the spin current  due to scattering is incorrect.

\bibitem{Rashba}
E.I. Rashba, cond-mat/0311110,  cond-mat/0404723.


\bibitem{Wigner}
It is a Wigner density matrix since we assume that the translation motion of the particles is the classical one. Yet, the spin dynamics is described exactly.


\bibitem{Koshelev}
A.E. Koshelev, V.Ya. Kravchenko, and D.E. Khmel'nitskii, Sov. Phys. Solid State {\bf 30}, 246 (1988).

\bibitem{Dyakonov2}
This is equivalent to calculating $q_{yz}$ through the following formula 
$q_{yz} \propto eE \int d^2{\bf p} (\partial/\partial p_x) S^{(0)}_y({\bf p})$, where $S^{(0)}_y({\bf p})= (\Omega_y({\bf p})/2 \Omega) (f^{(0)}_+ (p)- f^{(0)}_-(p))$ is the equilibrium value of the $S_y$ spin component. This integral is zero. (M.I. D'yakonov, private communication). 

\bibitem{Halperin}
E.G. Mishchenko, A.V. Shytov, and B.I. Halperin, 
 cond-mat/0406730.


\end{references}
\end{document}